\documentclass[journal=jacsat,manuscript=article]{achemso}

\usepackage[version=3]{mhchem} 

\author{Kamal Choudhary}
\email{kamal.choudhary@nist.gov}
\affiliation[National Institute of Standards and Technology]
{Materials Science and Engineering Division, National Institute of Standards and Technology, Gaithersburg, 20899, MD, USA.}
\alsoaffiliation[Theiss Research]
{Theiss Research, La Jolla, 92037, CA, USA.}

\author{Bobby G. Sumpter}
\affiliation[Oak Ridge National Laboratory]
{Center for Nanophase Materials Sciences, Oak Ridge National Laboratory, Oak Ridge, TN 37831, USA.}
\title
  {A Deep-learning Model for Fast Prediction of Vacancy Formation in Diverse Materials}

\abbreviations{IR,NMR,UV}

\begin{document}







\begin{abstract}
The presence of point defects such as vacancies plays an important role in material design. Here, we demonstrate that a graph neural network (GNN) model trained only on perfect materials can also be used to predict vacancy formation energies ($E_{vac}$) of defect structures without the need for additional training data. Such GNN-based predictions are considerably faster than density functional theory (DFT) calculations with reasonable accuracy and show the potential that GNNs are able to capture a functional form for energy predictions. To test this strategy, we developed a DFT dataset of 508 $E_{vac}$ consisting of 3D elemental solids, alloys, oxides, nitrides, and 2D monolayer materials. We analyzed and discussed the applicability of such direct and fast predictions. We applied the model to predict 192494 $E_{vac}$ for 55723 materials in the JARVIS-DFT database.
\end{abstract}


Defects play an important role in our pursuit to engineer performance of a material. Vacancies are a type of defects which are ubiquitous and their presence can significantly alter catalytic, electronic, optoelectronic, electrochemical, diffusion, and neuromorphic properties\cite{hinuma2018density,emery2016high,watkins2011large,emery2017high,yu2014vacancy,shang2016comprehensive,yan2019vacancy,parmar2018zn,ita1997effects,huh2020memristors}. Experimentally vacancy formation energies can be determined using positron annihilation experiments \cite{mckee1972vacancy}. Theoretically, they can be  computed using classical force-field (FF) \cite{choudhary2018high,huygh2014development} and density functional theory (DFT) \cite{medasani2015vacancy,ding2015pydii,goyal2020dopability,goyal2017computational,emery2016high} calculations. However, such computation can be very computationally expensive and non-generalizable for DFT  and FF based calculations respectively. 

Recently, machine learning techniques have been proposed as a faster method for predicting defect energetics, but so far they still require time-consuming defect data generation for model training and limit the applicability and generalizability of the defect energetic predictions \cite{cheng2020vacancy,arrigoni2021evolutionary,manzoor2021machine,frey2020machine,mannodi2022universal,sharma2020machine}. Especially, graph neural network based deep-learning models \cite{choudhary2022recent,choudhary2022graph,fung2021benchmarking,fung2021machine} have become very popular for predicting materials properties and have been used for several bulk property predictions and their applicability needs to be tested for defect property predictions. Two key ingredients needed for accomplishing this task are: 1) a pretrained deep-learning model that can directly predict the total energy of perfect and defect structures, 2) a test DFT dataset of vacancy formation energies on which the DL model could be applied.

In this work, we demonstrate that the atomistic line graph neural network (ALIGNN) \cite{choudhary2021atomistic} based total energy prediction model (trained on the JARVIS-DFT \cite{choudhary2020joint} OptB88vdW energy per atom data for perfect bulk materials \cite{choudhary2020joint}) can be directly used to predict vacancy formation energy of an arbitrary material with reasonable accuracy without requiring additional training data. The performance in terms of mean absolute error for the energy per atom model was reported as 0.037 eV in ref. \cite{choudhary2021atomistic}. Note that we do not train any machine learning/deep learning model in this work for defects and just used the model parameters for energy prediction that was developed and shared publicly in ref.\cite{choudhary2021atomistic}.

Developing a vacancy formation energy dataset can be extremely time-consuming and depends on several computational setup parameters such as supercell-size, choice of k-points, considering neutral vs. charged defects and selection of appropriate chemical potentials. For testing the strategy adopted in this work, we generated a DFT dataset of 508 entries with charge neutral defects using a high-throughput approach. The dataset consists of elemental solids, oxide, alloy and 2D materials. In addition to predicting the vacancy formation energies, we analyze the trends, strengths and limitations of such predictions. Lastly, we used this strategy to develop a database of vacancy formation energies for all the materials in the JARVIS-DFT database. The deep-learning model, the DFT dataset and the workflow are made publicly available through the JARVIS (Joint Automated Repository for Various Integrated Simulations) infrastructure \cite{choudhary2020joint}. 




First, we discuss the generation of vacancy formation energy dataset that is used for testing the deep-learning model. We obtained stable elemental solids, binary alloys, oxides and 2D materials from the JARVIS-DFT dataset. We used at least 8 $\AA$ lattice parameter constraints in x,y,z directions to build the supercell. We removed an atom with a unique Wyckoff position to generate the vacancy structure using the JARVIS-Tools package (\url {https://github.com/usnistgov/jarvis}). The defect structures were then subjected to energy minimization using Optb88vdW functional \cite{klimevs2009chemical} and projected augmented wave formalism \cite{blochl1994projector} in Vienna Ab initio Simulation Package (VASP) package \cite{kresse1996efficient,kresse1996efficiency}. Please note commercial software is identified to specify procedures. Such identification does not imply recommendation by National Institute of Standards and Technology (NIST). We used the converged k-point and cut-off from the JARVIS-DFT dataset based on total energy convergence \cite{choudhary2019convergence}. We used an energy convergence of $10^{-6}$ eV for energy convergence during the self-consistent cycle. Currently, we have 508 entries for the vacancies and the dataset is still growing.

For the deep-learning predictions, we used the recently developed atomistic line graph neural network (ALIGNN) \cite{choudhary2021atomistic}, which is publicly available at \url{https://github.com/usnistgov/alignn}. ALIGNN has been used to train fast and accurate models for more than 65 properties of solids and molecules with high accuracy\cite{choudhary2021atomistic,choudhary2022graph,choudhary2022designing,kaundinya2022prediction}.
In ALIGNN, a crystal structure is represented as a graph using atomic elements as nodes and atomic bonds as edges. Each node in the atomistic graph is assigned 9 input node features based on its atomic species:
electronegativity, group number, covalent radius, valence electrons, first ionization energy, electron affinity, block and atomic volume. The inter-atomic bond distances are used as edge features with radial basis function up to 8 $\textrm{\AA}$ cut-off and a 12-nearest-neighbor ($N$). This atomistic graph is then used for constructing the corresponding line graph using interatomic bond-distances as nodes and bond-angles as edge features. ALIGNN uses edge-gated graph convolution for updating nodes as well as edge features using a propagation function ($f$) for layer ($l$), atom features ($h$), and node ($i$), details of which can be found in Ref. \cite{choudhary2021atomistic}: 

\begin{equation} 
h_i^{(l+1)}=f(h_i^l{\{h_j^l}\}_{_i})
\end{equation}

Unlike many other conventional GNNs, ALIGNN uses bond-distances as well as bond-angles to distinguish atomic structures. The ALIGNN model is implemented in PyTorch \cite{paszke2019pytorch} and deep graph library (DGL) \cite{wang2019deep}. A model to predict energy per atom was developed in Ref\cite{choudhary2021atomistic} which will be used as an energy predictor for both perfect and defect structures in this work.

\begin{figure}[hbt!]
    \centering
    \includegraphics[trim={0. 0cm 0 0cm},clip,width=1\textwidth]{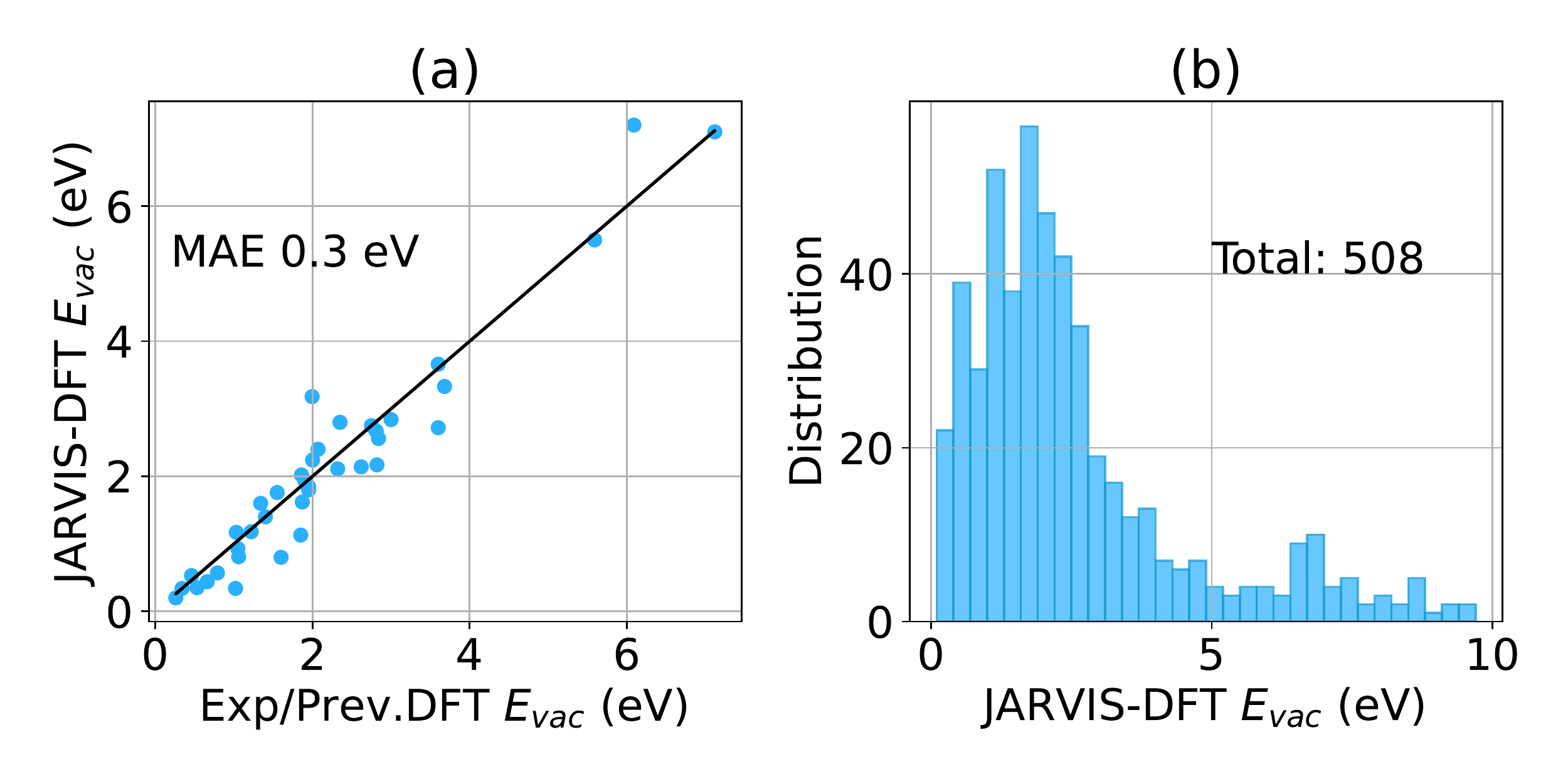}
    \caption{ Analysis of vacancy formation energy dataset generated in this work. a) comparison of a subset of vacancy formation energy with respect to available experimental and previous DFT calculations from literature. Our dataset agrees very well with previously reported values. b) Data distribution of all the vacancy formation energy values. Most of the values are below 3 eV.}
\end{figure}
In Fig. 1 we analyzed the DFT database for vacancy formation energies developed in this work. We used this dataset for testing purposes only. Although there have been several studies in generating vacancy formation energy dataset, a fully atomistic dataset consistent with bulk and vacancy energetics information is not available to our knowledge. Hence, we generated a DFT dataset for vacancies consisting of a wide variety of material classes such as elemental solids, 2D materials, oxides, and metallic alloys. We visualize the defect formation energies of materials in Fig. 1.
As mentioned above, we only considered the charge neutral vacancies within a finite 8 $\AA$ cell size with OptB88vdW functional. The vacancy formation energy was calculated as 
\begin{equation}
E_{vacancy}=E_{defect}-E_{perfect}+\mu
\end{equation}
where, $E_{vacancy}$ is the vacancy formation energy, $E_{defect}$ is the energy of the defect structure with an atom missing, $E_{perfect}$ is the energy of the perfect structure, $\mu$ is the chemical potential used as energy per atom of the most stable structure of an element. The chemical potentials used in this work are provided in the supplementary information.

\begin{figure}[hbt!]
    \centering
    \includegraphics[trim={0. 0cm 0 0cm},clip,width=1\textwidth]{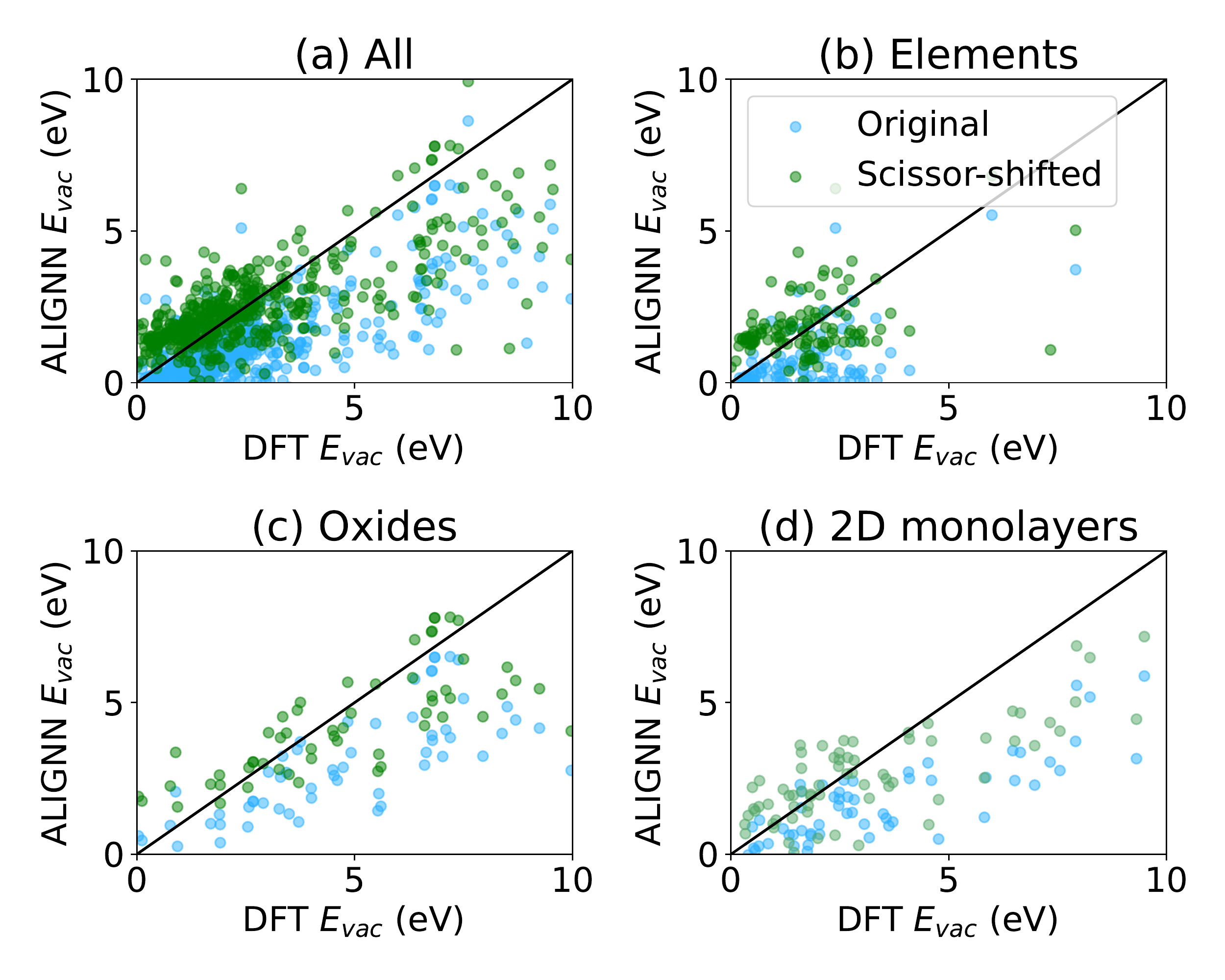}
    \caption{ Comparison of DFT and ALIGNN based vacancy formation energy predictions with (Scissor-shift) and without (Original) correction. In Fig. a, we show the performance on the entire DFT dataset generated in this work. Fig.b,c,d show the comparisons for elemental solids, oxides and 2D monolayers respectively.}
\end{figure}

Currently, the vacancy formation energy dataset consists of 508 entries. We compare a subset of this dataset with available data from previous experimental and DFT-studies \cite{medasani2015vacancy,pham2015oxygen,kumar2015charge,choudhary2015charge,guo2015chalcogen,wu2020first} in Fig. 1a. We find an excellent agreement between our dataset and that from literature with a mean absolute error (MAE) of 0.3 eV. In Fig. 1b, we show the histogram of all the vacancy formation energy data. We find that most of the vacancy formation energy data lie below 3 eV. Depending on the type of engineering applications, either a high or low $E_{vac}$ could be desirable. 
\begin{figure}[hbt!]
    \centering
    \includegraphics[trim={0. 0cm 0 0cm},clip,width=1\textwidth]{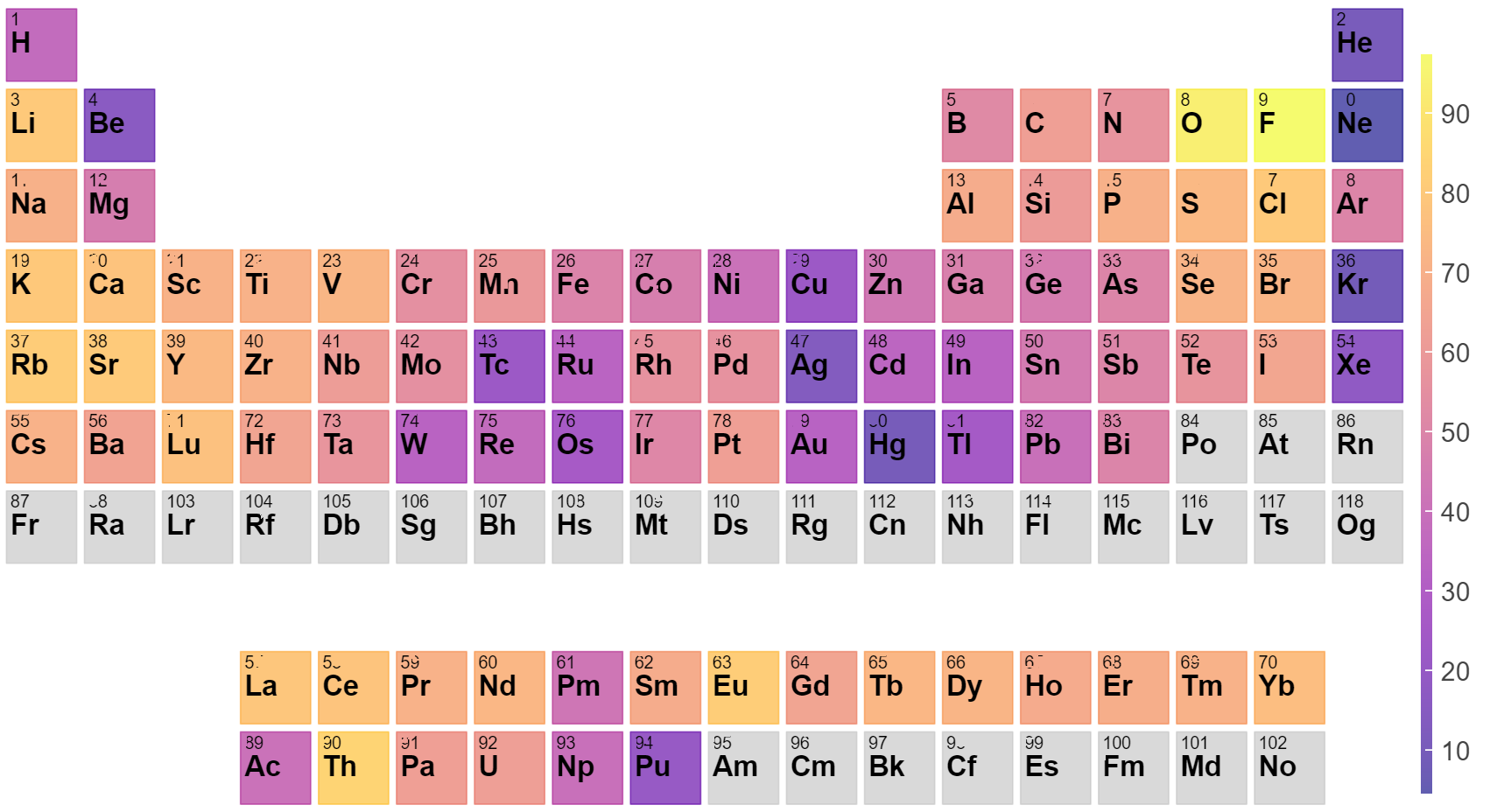}
    \caption{ Periodic table trend of 192494 vacancy formation energies predicted from scissor-shifted ALIGNN based predictions. We visualize the probability that compounds with vacancy of a given element have vacancy formation energy more than 2 eV.}
\end{figure}
\begin{table}
\caption{Analysis and comparison of ALIGNN based vacancy formation energy with respect to DFT dataset. The mean absolute errors (MAE) are calculated for all the values as well as elemental solid, oxide and 2D-monolayer subsets. The corresponding number of entries (Count) are also provided. As the ALIGNN based vacancy formation energies could be underestimated we apply scissor-shift and the improvement (\% $\Delta$) of predictions with (Scissor-shifted) and without scissor shift (Original) are shown. }
\begin{minipage}{400pt}
\begin{tabular}{|c|c|c|c|c|}
Sets & Count & Original MAE  & Scissor-shift MAE & \% $\Delta$\\
All& 508 & 1.51 & 1.0&  33.8\\
Elements& 117 & 1.53 & 1.2& 21.6  \\
Oxides& 57 & 2.3 & 1.4& 30.4 \\
2D-mono& 68 & 1.9 & 1.2&36.8  \\
\end{tabular}
\end{minipage}
\end{table}

\begin{figure}[hbt!]
    \centering
    \includegraphics[trim={0. 0cm 0 0cm},clip,width=1\textwidth]{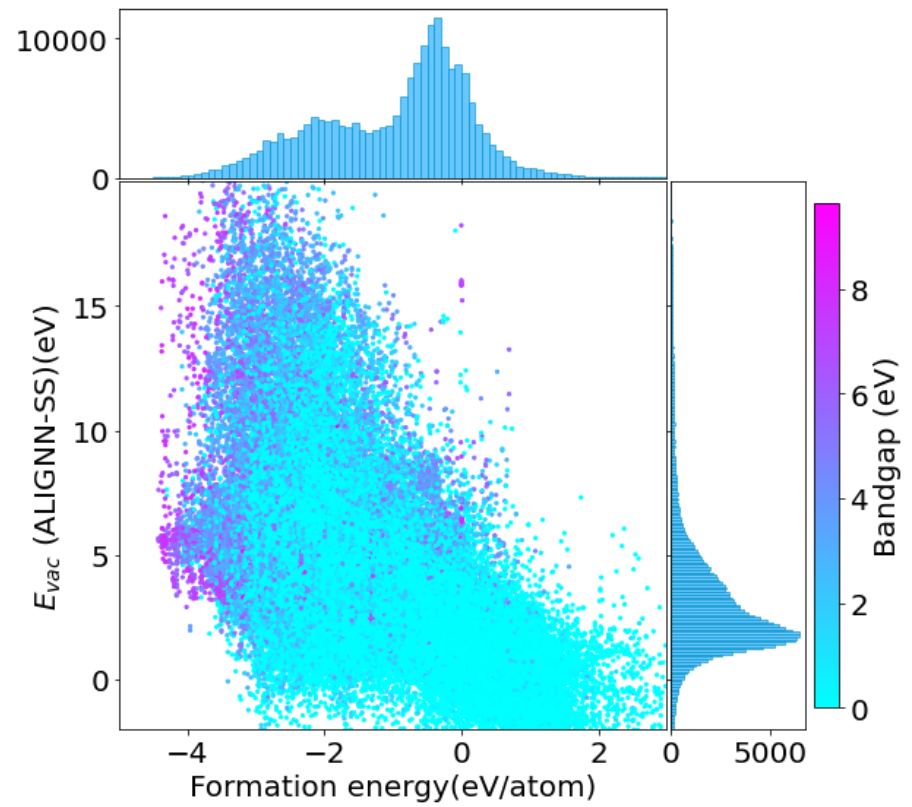}
    \caption{Figure shows ALIGNN vacancy formation energy with scissor-shift (SS) dataset (192494 entires) against OptB88vdW based formation energy available in the JARVIS-DFT database and color coded with corresponding OptB88vdW bandgaps.}
\end{figure}

Next, we used ALIGNN based pretrained total energy per atom model trained on the JARVIS-DFT dataset for predicting defect energy and perfect energy required for vacancy formation energy following Eq. 2. This model was trained using bulk energies for 55723 solids \cite{choudhary2021atomistic}. The defect structures were generated by deleting an atom with a unique Wyckoff position without optimizing the atomic positions of other atoms in the defect structures. We used the same chemical potential for elements from the JARVIS-DFT as given in the supporting information. The comparison of ALIGNN based prediction with respect to DFT data is shown with blue dots in Fig. 2a. Interestingly, we observed that there is a noticeable correlation between the ALIGNN direct predictions and DFT data with a mean absolute error of 1.51 eV. However, we found that the ALIGNN based predictions were usually underestimated. To circumvent this issue, we applied a scissor shift by adding 1.3 eV to all the ALIGNN based predictions represented by green dots. Using such a shift, we were able to lower the MAE to 1.0 eV i.e., leading to an overall 33.8 \% improvement. The value of 1.3 eV was chosen such that overall MAE is minimized. We noted that the previous reports on machine learning for vacancy formation energies resulted in mean absolute error values of 0.40 eV and 0.67 eV in ref. \cite{cheng2020vacancy} and ref. \cite{frey2020machine} respectively. These models were trained on specific material classes such as GeTe system and 2D materials, while the approach used here acts as a generalized application of the model with MAE closer to specific case studies. Additionally, an MAE of 1.0 eV is arguably low considering the range of the vacancy formation energies can be upto 10 eV.

To further analyze the predictions for different types of materials, we compared the DFT and ALIGNN predictions for elemental solids, oxides and 2D monolayers in Fig. 2b, Fig. 2c and Fig. 2d respectively. Corresponding original ALIGNN prediction based MAE and that with scissor-shifted values are shown in Table 1. We found that the ALIGNN based models perform well for 2D monolayer and oxide materials compared to elemental solids and alloys. This behavior can be explained based on the fact that GNN architectures usually perform message passing locally and may work well for insulting materials with fewer bonds rather than for elemental solids and alloy systems which are usually metallic and have delocalized electronic structures leading to a higher number of interatomic bonds. 

To analyze the effect of training dataset size of energy per atom ALIGNN model on prediction of vacancy formation energies, we trained models for 20 \%, 40 \%, 60 \%, 80 \% and 100 \% of total energy per atom training data for perfect materials resulting in mean absolute errors of 1.20 eV, 1.22 eV, 1.11 eV, 1.05 eV and 1.0 eV for vacancy formation dataset respectively. Therefore, one possible way to further improve the model would be to include more perfect structures that can have a varied number of bonds for various systems. As the JARVIS-DFT dataset for perfect structures is still growing, we believe the defect property prediction models should further improve in the future.



Now, we applied the above strategy to predict the vacancy formation energies of all the 55723 materials in the JARVIS-DFT leading to 192494 vacancy formation energies. In Fig. 3, we visualize the probability that compounds with the vacancy of a given element have vacancy formation energy of more than 2 eV as a threshold value. Interestingly, we found that C,N,O,F are some of the common elements with high vacancy formation energies. Furthermore, we plot the ALIGNN based vacancy formation energy dataset against OptB88vdW based formation energy available in the JARVIS-DFT database and color code with the corresponding OptB88vdW bandgaps in Fig. 4. We also provided the histogram distribution of the formation energy per atom and vacancy formation energy per atom. Interestingly, we found that high vacancy formation energies were favored by lower formation energies per atom and high electronic bandgaps. The vacancy formation energy histogram showed a high peak around 2 eV which was similar to that observed in Fig. 1b. Such analysis helped us understand the entire dataset in a nutshell and can further help in understanding defect energetics for materials design for instance finding materials with high dopability, water splitting etc.


In summary, we have developed a diverse dataset of vacancy formation energies using density functional theory and demonstrate that the total energy per atom model using ALIGNN can be directly used for predicting vacancy formation energies without the need for additional training data. The defect database can be highly useful for materials science community as there is a strong demand for a benchmark dataset. We have discussed the assumptions used in this work such as excluding charge defects and using finite cell sizes. Also, our work has shown how a state of the art model performs on unseen data without additional data. We have applied a heuristic scissor-shift of energy that further improves the accuracy. Using the current strategy, we have predicted vacancy formation energies of around 55723 compounds with 192494 entries leading to the largest dataset of defect properties, which could have been very expensive from DFT calculations. We provide data from this work as well as machine learning models to help accelerate the design of new materials. Our work has proven that GNN models can not only be useful for perfect materials but also defect systems.

\clearpage
\begin{acknowledgement}

K.C. thanks the National Institute of Standards and Technology for funding, computational, and
data-management resources. Contributions from K.C. were supported by the financial assistance award 70NANB19H117 from the U.S. Department of Commerce, National Institute of Standards and Technology. DFT calculations were also conducted at the Center for Nanophase Materials Sciences, a US Department of Energy Office of Science User Facility. 

\end{acknowledgement}

\begin{suppinfo}

Experimental and previous DFT data for benchmarking, and chemical potential of elemental solids are provided in the supporting information.

\end{suppinfo}

\bibliography{achemso-demo}

\providecommand{\latin}[1]{#1}
\makeatletter
\providecommand{\doi}
  {\begingroup\let\do\@makeother\dospecials
  \catcode`\{=1 \catcode`\}=2 \doi@aux}
\providecommand{\doi@aux}[1]{\endgroup\texttt{#1}}
\makeatother
\providecommand*\mcitethebibliography{\thebibliography}
\csname @ifundefined\endcsname{endmcitethebibliography}
  {\let\endmcitethebibliography\endthebibliography}{}
\begin{mcitethebibliography}{44}
\providecommand*\natexlab[1]{#1}
\providecommand*\mciteSetBstSublistMode[1]{}
\providecommand*\mciteSetBstMaxWidthForm[2]{}
\providecommand*\mciteBstWouldAddEndPuncttrue
  {\def\EndOfBibitem{\unskip.}}
\providecommand*\mciteBstWouldAddEndPunctfalse
  {\let\EndOfBibitem\relax}
\providecommand*\mciteSetBstMidEndSepPunct[3]{}
\providecommand*\mciteSetBstSublistLabelBeginEnd[3]{}
\providecommand*\EndOfBibitem{}
\mciteSetBstSublistMode{f}
\mciteSetBstMaxWidthForm{subitem}{(\alph{mcitesubitemcount})}
\mciteSetBstSublistLabelBeginEnd
  {\mcitemaxwidthsubitemform\space}
  {\relax}
  {\relax}

\bibitem[Hinuma \latin{et~al.}(2018)Hinuma, Toyao, Kamachi, Maeno, Takakusagi,
  Furukawa, Takigawa, and Shimizu]{hinuma2018density}
Hinuma,~Y.; Toyao,~T.; Kamachi,~T.; Maeno,~Z.; Takakusagi,~S.; Furukawa,~S.;
  Takigawa,~I.; Shimizu,~K.-i. Density functional theory calculations of oxygen
  vacancy formation and subsequent molecular adsorption on oxide surfaces.
  \emph{The Journal of Physical Chemistry C} \textbf{2018}, \emph{122},
  29435--29444\relax
\mciteBstWouldAddEndPuncttrue
\mciteSetBstMidEndSepPunct{\mcitedefaultmidpunct}
{\mcitedefaultendpunct}{\mcitedefaultseppunct}\relax
\EndOfBibitem
\bibitem[Emery \latin{et~al.}(2016)Emery, Saal, Kirklin, Hegde, and
  Wolverton]{emery2016high}
Emery,~A.~A.; Saal,~J.~E.; Kirklin,~S.; Hegde,~V.~I.; Wolverton,~C.
  High-throughput computational screening of perovskites for thermochemical
  water splitting applications. \emph{Chemistry of Materials} \textbf{2016},
  \emph{28}, 5621--5634\relax
\mciteBstWouldAddEndPuncttrue
\mciteSetBstMidEndSepPunct{\mcitedefaultmidpunct}
{\mcitedefaultendpunct}{\mcitedefaultseppunct}\relax
\EndOfBibitem
\bibitem[Watkins \latin{et~al.}(2011)Watkins, Pan, Wang, Michaelides,
  VandeVondele, and Slater]{watkins2011large}
Watkins,~M.; Pan,~D.; Wang,~E.~G.; Michaelides,~A.; VandeVondele,~J.;
  Slater,~B. Large variation of vacancy formation energies in the surface of
  crystalline ice. \emph{Nature materials} \textbf{2011}, \emph{10},
  794--798\relax
\mciteBstWouldAddEndPuncttrue
\mciteSetBstMidEndSepPunct{\mcitedefaultmidpunct}
{\mcitedefaultendpunct}{\mcitedefaultseppunct}\relax
\EndOfBibitem
\bibitem[Emery and Wolverton(2017)Emery, and Wolverton]{emery2017high}
Emery,~A.~A.; Wolverton,~C. High-throughput DFT calculations of formation
  energy, stability and oxygen vacancy formation energy of ABO3 perovskites.
  \emph{Scientific data} \textbf{2017}, \emph{4}, 1--10\relax
\mciteBstWouldAddEndPuncttrue
\mciteSetBstMidEndSepPunct{\mcitedefaultmidpunct}
{\mcitedefaultendpunct}{\mcitedefaultseppunct}\relax
\EndOfBibitem
\bibitem[Yu \latin{et~al.}(2014)Yu, Zhan, Rong, Liu, Li, and
  Liu]{yu2014vacancy}
Yu,~X.; Zhan,~Z.; Rong,~J.; Liu,~Z.; Li,~L.; Liu,~J. Vacancy formation energy
  and size effects. \emph{Chemical Physics Letters} \textbf{2014}, \emph{600},
  43--45\relax
\mciteBstWouldAddEndPuncttrue
\mciteSetBstMidEndSepPunct{\mcitedefaultmidpunct}
{\mcitedefaultendpunct}{\mcitedefaultseppunct}\relax
\EndOfBibitem
\bibitem[Shang \latin{et~al.}(2016)Shang, Zhou, Wang, Ross, Liu, Hu, Fang,
  Wang, and Liu]{shang2016comprehensive}
Shang,~S.-L.; Zhou,~B.-C.; Wang,~W.~Y.; Ross,~A.~J.; Liu,~X.~L.; Hu,~Y.-J.;
  Fang,~H.-Z.; Wang,~Y.; Liu,~Z.-K. A comprehensive first-principles study of
  pure elements: Vacancy formation and migration energies and self-diffusion
  coefficients. \emph{Acta Materialia} \textbf{2016}, \emph{109},
  128--141\relax
\mciteBstWouldAddEndPuncttrue
\mciteSetBstMidEndSepPunct{\mcitedefaultmidpunct}
{\mcitedefaultendpunct}{\mcitedefaultseppunct}\relax
\EndOfBibitem
\bibitem[Yan \latin{et~al.}(2019)Yan, Zhao, Chen, Zhao, Zhou, Wang, Wang,
  Zhang, Li, Xiao, \latin{et~al.} others]{yan2019vacancy}
Yan,~X.; Zhao,~Q.; Chen,~A.~P.; Zhao,~J.; Zhou,~Z.; Wang,~J.; Wang,~H.;
  Zhang,~L.; Li,~X.; Xiao,~Z., \latin{et~al.}  Vacancy-induced synaptic
  behavior in 2D WS2 nanosheet--based memristor for low-power neuromorphic
  computing. \emph{Small} \textbf{2019}, \emph{15}, 1901423\relax
\mciteBstWouldAddEndPuncttrue
\mciteSetBstMidEndSepPunct{\mcitedefaultmidpunct}
{\mcitedefaultendpunct}{\mcitedefaultseppunct}\relax
\EndOfBibitem
\bibitem[Parmar \latin{et~al.}(2018)Parmar, Boatner, Lynn, and
  Choi]{parmar2018zn}
Parmar,~N.~S.; Boatner,~L.~A.; Lynn,~K.~G.; Choi,~J.-W. Zn vacancy formation
  energy and diffusion coefficient of CVT ZnO crystals in the sub-surface
  micron region. \emph{Scientific reports} \textbf{2018}, \emph{8}, 1--8\relax
\mciteBstWouldAddEndPuncttrue
\mciteSetBstMidEndSepPunct{\mcitedefaultmidpunct}
{\mcitedefaultendpunct}{\mcitedefaultseppunct}\relax
\EndOfBibitem
\bibitem[Ita and Cohen(1997)Ita, and Cohen]{ita1997effects}
Ita,~J.; Cohen,~R.~E. Effects of pressure on diffusion and vacancy formation in
  MgO from nonempirical free-energy integrations. \emph{Physical Review
  Letters} \textbf{1997}, \emph{79}, 3198\relax
\mciteBstWouldAddEndPuncttrue
\mciteSetBstMidEndSepPunct{\mcitedefaultmidpunct}
{\mcitedefaultendpunct}{\mcitedefaultseppunct}\relax
\EndOfBibitem
\bibitem[Huh \latin{et~al.}(2020)Huh, Lee, and Lee]{huh2020memristors}
Huh,~W.; Lee,~D.; Lee,~C.-H. Memristors based on 2D materials as an artificial
  synapse for neuromorphic electronics. \emph{Advanced Materials}
  \textbf{2020}, \emph{32}, 2002092\relax
\mciteBstWouldAddEndPuncttrue
\mciteSetBstMidEndSepPunct{\mcitedefaultmidpunct}
{\mcitedefaultendpunct}{\mcitedefaultseppunct}\relax
\EndOfBibitem
\bibitem[McKee \latin{et~al.}(1972)McKee, Triftsh{\"a}user, and
  Stewart]{mckee1972vacancy}
McKee,~B.; Triftsh{\"a}user,~W.; Stewart,~A. Vacancy-formation energies in
  metals from positron annihilation. \emph{Physical Review Letters}
  \textbf{1972}, \emph{28}, 358\relax
\mciteBstWouldAddEndPuncttrue
\mciteSetBstMidEndSepPunct{\mcitedefaultmidpunct}
{\mcitedefaultendpunct}{\mcitedefaultseppunct}\relax
\EndOfBibitem
\bibitem[Choudhary \latin{et~al.}(2018)Choudhary, Biacchi, Ghosh, Hale, Walker,
  and Tavazza]{choudhary2018high}
Choudhary,~K.; Biacchi,~A.~J.; Ghosh,~S.; Hale,~L.; Walker,~A. R.~H.;
  Tavazza,~F. High-throughput assessment of vacancy formation and surface
  energies of materials using classical force-fields. \emph{Journal of Physics:
  Condensed Matter} \textbf{2018}, \emph{30}, 395901\relax
\mciteBstWouldAddEndPuncttrue
\mciteSetBstMidEndSepPunct{\mcitedefaultmidpunct}
{\mcitedefaultendpunct}{\mcitedefaultseppunct}\relax
\EndOfBibitem
\bibitem[Huygh \latin{et~al.}(2014)Huygh, Bogaerts, Van~Duin, and
  Neyts]{huygh2014development}
Huygh,~S.; Bogaerts,~A.; Van~Duin,~A.~C.; Neyts,~E.~C. Development of a ReaxFF
  reactive force field for intrinsic point defects in titanium dioxide.
  \emph{Computational materials science} \textbf{2014}, \emph{95},
  579--591\relax
\mciteBstWouldAddEndPuncttrue
\mciteSetBstMidEndSepPunct{\mcitedefaultmidpunct}
{\mcitedefaultendpunct}{\mcitedefaultseppunct}\relax
\EndOfBibitem
\bibitem[Medasani \latin{et~al.}(2015)Medasani, Haranczyk, Canning, and
  Asta]{medasani2015vacancy}
Medasani,~B.; Haranczyk,~M.; Canning,~A.; Asta,~M. Vacancy formation energies
  in metals: A comparison of MetaGGA with LDA and GGA exchange--correlation
  functionals. \emph{Computational Materials Science} \textbf{2015},
  \emph{101}, 96--107\relax
\mciteBstWouldAddEndPuncttrue
\mciteSetBstMidEndSepPunct{\mcitedefaultmidpunct}
{\mcitedefaultendpunct}{\mcitedefaultseppunct}\relax
\EndOfBibitem
\bibitem[Ding \latin{et~al.}(2015)Ding, Medasani, Chen, Persson, Haranczyk, and
  Asta]{ding2015pydii}
Ding,~H.; Medasani,~B.; Chen,~W.; Persson,~K.~A.; Haranczyk,~M.; Asta,~M.
  PyDII: a Python framework for computing equilibrium intrinsic point defect
  concentrations and extrinsic solute site preferences in intermetallic
  compounds. \emph{Computer Physics Communications} \textbf{2015}, \emph{193},
  118--123\relax
\mciteBstWouldAddEndPuncttrue
\mciteSetBstMidEndSepPunct{\mcitedefaultmidpunct}
{\mcitedefaultendpunct}{\mcitedefaultseppunct}\relax
\EndOfBibitem
\bibitem[Goyal \latin{et~al.}(2020)Goyal, Gorai, Anand, Toberer, Snyder, and
  Stevanovic]{goyal2020dopability}
Goyal,~A.; Gorai,~P.; Anand,~S.; Toberer,~E.~S.; Snyder,~G.~J.; Stevanovic,~V.
  On the dopability of semiconductors and governing material properties.
  \emph{Chemistry of Materials} \textbf{2020}, \emph{32}, 4467--4480\relax
\mciteBstWouldAddEndPuncttrue
\mciteSetBstMidEndSepPunct{\mcitedefaultmidpunct}
{\mcitedefaultendpunct}{\mcitedefaultseppunct}\relax
\EndOfBibitem
\bibitem[Goyal \latin{et~al.}(2017)Goyal, Gorai, Peng, Lany, and
  Stevanovi{\'c}]{goyal2017computational}
Goyal,~A.; Gorai,~P.; Peng,~H.; Lany,~S.; Stevanovi{\'c},~V. A computational
  framework for automation of point defect calculations. \emph{Computational
  Materials Science} \textbf{2017}, \emph{130}, 1--9\relax
\mciteBstWouldAddEndPuncttrue
\mciteSetBstMidEndSepPunct{\mcitedefaultmidpunct}
{\mcitedefaultendpunct}{\mcitedefaultseppunct}\relax
\EndOfBibitem
\bibitem[Cheng \latin{et~al.}(2020)Cheng, Zhu, Wang, Zhou, Elliott, and
  Sun]{cheng2020vacancy}
Cheng,~Y.; Zhu,~L.; Wang,~G.; Zhou,~J.; Elliott,~S.~R.; Sun,~Z. Vacancy
  formation energy and its connection with bonding environment in solid: A
  high-throughput calculation and machine learning study. \emph{Computational
  Materials Science} \textbf{2020}, \emph{183}, 109803\relax
\mciteBstWouldAddEndPuncttrue
\mciteSetBstMidEndSepPunct{\mcitedefaultmidpunct}
{\mcitedefaultendpunct}{\mcitedefaultseppunct}\relax
\EndOfBibitem
\bibitem[Arrigoni and Madsen(2021)Arrigoni, and
  Madsen]{arrigoni2021evolutionary}
Arrigoni,~M.; Madsen,~G.~K. Evolutionary computing and machine learning for
  discovering of low-energy defect configurations. \emph{npj Computational
  Materials} \textbf{2021}, \emph{7}, 1--13\relax
\mciteBstWouldAddEndPuncttrue
\mciteSetBstMidEndSepPunct{\mcitedefaultmidpunct}
{\mcitedefaultendpunct}{\mcitedefaultseppunct}\relax
\EndOfBibitem
\bibitem[Manzoor \latin{et~al.}(2021)Manzoor, Arora, Jerome, Linton, Norman,
  and Aidhy]{manzoor2021machine}
Manzoor,~A.; Arora,~G.; Jerome,~B.; Linton,~N.; Norman,~B.; Aidhy,~D.~S.
  Machine Learning Based Methodology to Predict Point Defect Energies in
  Multi-Principal Element Alloys. \emph{Frontiers in Materials} \textbf{2021},
  \emph{8}, 129\relax
\mciteBstWouldAddEndPuncttrue
\mciteSetBstMidEndSepPunct{\mcitedefaultmidpunct}
{\mcitedefaultendpunct}{\mcitedefaultseppunct}\relax
\EndOfBibitem
\bibitem[Frey \latin{et~al.}(2020)Frey, Akinwande, Jariwala, and
  Shenoy]{frey2020machine}
Frey,~N.~C.; Akinwande,~D.; Jariwala,~D.; Shenoy,~V.~B. Machine
  learning-enabled design of point defects in 2d materials for quantum and
  neuromorphic information processing. \emph{ACS nano} \textbf{2020},
  \emph{14}, 13406--13417\relax
\mciteBstWouldAddEndPuncttrue
\mciteSetBstMidEndSepPunct{\mcitedefaultmidpunct}
{\mcitedefaultendpunct}{\mcitedefaultseppunct}\relax
\EndOfBibitem
\bibitem[Mannodi-Kanakkithodi \latin{et~al.}(2022)Mannodi-Kanakkithodi, Xiang,
  Jacoby, Biegaj, Dunham, Gamelin, and Chan]{mannodi2022universal}
Mannodi-Kanakkithodi,~A.; Xiang,~X.; Jacoby,~L.; Biegaj,~R.; Dunham,~S.~T.;
  Gamelin,~D.~R.; Chan,~M.~K. Universal machine learning framework for defect
  predictions in zinc blende semiconductors. \emph{Patterns} \textbf{2022},
  \emph{3}, 100450\relax
\mciteBstWouldAddEndPuncttrue
\mciteSetBstMidEndSepPunct{\mcitedefaultmidpunct}
{\mcitedefaultendpunct}{\mcitedefaultseppunct}\relax
\EndOfBibitem
\bibitem[Sharma \latin{et~al.}(2020)Sharma, Kumar, Dev, and
  Pilania]{sharma2020machine}
Sharma,~V.; Kumar,~P.; Dev,~P.; Pilania,~G. Machine learning substitutional
  defect formation energies in ABO3 perovskites. \emph{Journal of Applied
  Physics} \textbf{2020}, \emph{128}, 034902\relax
\mciteBstWouldAddEndPuncttrue
\mciteSetBstMidEndSepPunct{\mcitedefaultmidpunct}
{\mcitedefaultendpunct}{\mcitedefaultseppunct}\relax
\EndOfBibitem
\bibitem[Choudhary \latin{et~al.}(2022)Choudhary, DeCost, Chen, Jain, Tavazza,
  Cohn, Park, Choudhary, Agrawal, Billinge, \latin{et~al.}
  others]{choudhary2022recent}
Choudhary,~K.; DeCost,~B.; Chen,~C.; Jain,~A.; Tavazza,~F.; Cohn,~R.;
  Park,~C.~W.; Choudhary,~A.; Agrawal,~A.; Billinge,~S.~J., \latin{et~al.}
  Recent advances and applications of deep learning methods in materials
  science. \emph{npj Computational Materials} \textbf{2022}, \emph{8},
  1--26\relax
\mciteBstWouldAddEndPuncttrue
\mciteSetBstMidEndSepPunct{\mcitedefaultmidpunct}
{\mcitedefaultendpunct}{\mcitedefaultseppunct}\relax
\EndOfBibitem
\bibitem[Choudhary \latin{et~al.}(2022)Choudhary, Yildirim, Siderius, Kusne,
  McDannald, and Ortiz-Montalvo]{choudhary2022graph}
Choudhary,~K.; Yildirim,~T.; Siderius,~D.~W.; Kusne,~A.~G.; McDannald,~A.;
  Ortiz-Montalvo,~D.~L. Graph neural network predictions of metal organic
  framework CO2 adsorption properties. \emph{Computational Materials Science}
  \textbf{2022}, \emph{210}, 111388\relax
\mciteBstWouldAddEndPuncttrue
\mciteSetBstMidEndSepPunct{\mcitedefaultmidpunct}
{\mcitedefaultendpunct}{\mcitedefaultseppunct}\relax
\EndOfBibitem
\bibitem[Fung \latin{et~al.}(2021)Fung, Zhang, Juarez, and
  Sumpter]{fung2021benchmarking}
Fung,~V.; Zhang,~J.; Juarez,~E.; Sumpter,~B.~G. Benchmarking graph neural
  networks for materials chemistry. \emph{npj Computational Materials}
  \textbf{2021}, \emph{7}, 1--8\relax
\mciteBstWouldAddEndPuncttrue
\mciteSetBstMidEndSepPunct{\mcitedefaultmidpunct}
{\mcitedefaultendpunct}{\mcitedefaultseppunct}\relax
\EndOfBibitem
\bibitem[Fung \latin{et~al.}(2021)Fung, Hu, Ganesh, and
  Sumpter]{fung2021machine}
Fung,~V.; Hu,~G.; Ganesh,~P.; Sumpter,~B.~G. Machine learned features from
  density of states for accurate adsorption energy prediction. \emph{Nature
  communications} \textbf{2021}, \emph{12}, 1--11\relax
\mciteBstWouldAddEndPuncttrue
\mciteSetBstMidEndSepPunct{\mcitedefaultmidpunct}
{\mcitedefaultendpunct}{\mcitedefaultseppunct}\relax
\EndOfBibitem
\bibitem[Choudhary and DeCost(2021)Choudhary, and
  DeCost]{choudhary2021atomistic}
Choudhary,~K.; DeCost,~B. Atomistic Line Graph Neural Network for improved
  materials property predictions. \emph{npj Computational Materials}
  \textbf{2021}, \emph{7}, 1--8\relax
\mciteBstWouldAddEndPuncttrue
\mciteSetBstMidEndSepPunct{\mcitedefaultmidpunct}
{\mcitedefaultendpunct}{\mcitedefaultseppunct}\relax
\EndOfBibitem
\bibitem[Choudhary \latin{et~al.}(2020)Choudhary, Garrity, Reid, DeCost,
  Biacchi, Hight~Walker, Trautt, Hattrick-Simpers, Kusne, Centrone,
  \latin{et~al.} others]{choudhary2020joint}
Choudhary,~K.; Garrity,~K.~F.; Reid,~A.~C.; DeCost,~B.; Biacchi,~A.~J.;
  Hight~Walker,~A.~R.; Trautt,~Z.; Hattrick-Simpers,~J.; Kusne,~A.~G.;
  Centrone,~A., \latin{et~al.}  The joint automated repository for various
  integrated simulations (JARVIS) for data-driven materials design. \emph{npj
  Computational Materials} \textbf{2020}, \emph{6}, 1--13\relax
\mciteBstWouldAddEndPuncttrue
\mciteSetBstMidEndSepPunct{\mcitedefaultmidpunct}
{\mcitedefaultendpunct}{\mcitedefaultseppunct}\relax
\EndOfBibitem
\bibitem[Klime{\v{s}} \latin{et~al.}(2009)Klime{\v{s}}, Bowler, and
  Michaelides]{klimevs2009chemical}
Klime{\v{s}},~J.; Bowler,~D.~R.; Michaelides,~A. Chemical accuracy for the van
  der Waals density functional. \emph{Journal of Physics: Condensed Matter}
  \textbf{2009}, \emph{22}, 022201\relax
\mciteBstWouldAddEndPuncttrue
\mciteSetBstMidEndSepPunct{\mcitedefaultmidpunct}
{\mcitedefaultendpunct}{\mcitedefaultseppunct}\relax
\EndOfBibitem
\bibitem[Bl{\"o}chl(1994)]{blochl1994projector}
Bl{\"o}chl,~P.~E. Projector augmented-wave method. \emph{Physical review B}
  \textbf{1994}, \emph{50}, 17953\relax
\mciteBstWouldAddEndPuncttrue
\mciteSetBstMidEndSepPunct{\mcitedefaultmidpunct}
{\mcitedefaultendpunct}{\mcitedefaultseppunct}\relax
\EndOfBibitem
\bibitem[Kresse and Furthm{\"u}ller(1996)Kresse, and
  Furthm{\"u}ller]{kresse1996efficient}
Kresse,~G.; Furthm{\"u}ller,~J. Efficient iterative schemes for ab initio
  total-energy calculations using a plane-wave basis set. \emph{Physical review
  B} \textbf{1996}, \emph{54}, 11169\relax
\mciteBstWouldAddEndPuncttrue
\mciteSetBstMidEndSepPunct{\mcitedefaultmidpunct}
{\mcitedefaultendpunct}{\mcitedefaultseppunct}\relax
\EndOfBibitem
\bibitem[Kresse and Furthm{\"u}ller(1996)Kresse, and
  Furthm{\"u}ller]{kresse1996efficiency}
Kresse,~G.; Furthm{\"u}ller,~J. Efficiency of ab-initio total energy
  calculations for metals and semiconductors using a plane-wave basis set.
  \emph{Computational materials science} \textbf{1996}, \emph{6}, 15--50\relax
\mciteBstWouldAddEndPuncttrue
\mciteSetBstMidEndSepPunct{\mcitedefaultmidpunct}
{\mcitedefaultendpunct}{\mcitedefaultseppunct}\relax
\EndOfBibitem
\bibitem[Choudhary and Tavazza(2019)Choudhary, and
  Tavazza]{choudhary2019convergence}
Choudhary,~K.; Tavazza,~F. Convergence and machine learning predictions of
  Monkhorst-Pack k-points and plane-wave cut-off in high-throughput DFT
  calculations. \emph{Computational materials science} \textbf{2019},
  \emph{161}, 300--308\relax
\mciteBstWouldAddEndPuncttrue
\mciteSetBstMidEndSepPunct{\mcitedefaultmidpunct}
{\mcitedefaultendpunct}{\mcitedefaultseppunct}\relax
\EndOfBibitem
\bibitem[Choudhary and Garrity(2022)Choudhary, and
  Garrity]{choudhary2022designing}
Choudhary,~K.; Garrity,~K. Designing High-Tc Superconductors with BCS-inspired
  Screening, Density Functional Theory and Deep-learning. \emph{arXiv preprint
  arXiv:2205.00060} \textbf{2022}, \relax
\mciteBstWouldAddEndPunctfalse
\mciteSetBstMidEndSepPunct{\mcitedefaultmidpunct}
{}{\mcitedefaultseppunct}\relax
\EndOfBibitem
\bibitem[Kaundinya \latin{et~al.}(2022)Kaundinya, Choudhary, and
  Kalidindi]{kaundinya2022prediction}
Kaundinya,~P.~R.; Choudhary,~K.; Kalidindi,~S.~R. Prediction of the electron
  density of states for crystalline compounds with Atomistic Line Graph Neural
  Networks (ALIGNN). \emph{arXiv preprint arXiv:2201.08348} \textbf{2022},
  \relax
\mciteBstWouldAddEndPunctfalse
\mciteSetBstMidEndSepPunct{\mcitedefaultmidpunct}
{}{\mcitedefaultseppunct}\relax
\EndOfBibitem
\bibitem[Paszke \latin{et~al.}(2019)Paszke, Gross, Massa, Lerer, Bradbury,
  Chanan, Killeen, Lin, Gimelshein, Antiga, \latin{et~al.}
  others]{paszke2019pytorch}
Paszke,~A.; Gross,~S.; Massa,~F.; Lerer,~A.; Bradbury,~J.; Chanan,~G.;
  Killeen,~T.; Lin,~Z.; Gimelshein,~N.; Antiga,~L., \latin{et~al.}  Pytorch: An
  imperative style, high-performance deep learning library. \emph{Advances in
  neural information processing systems} \textbf{2019}, \emph{32}\relax
\mciteBstWouldAddEndPuncttrue
\mciteSetBstMidEndSepPunct{\mcitedefaultmidpunct}
{\mcitedefaultendpunct}{\mcitedefaultseppunct}\relax
\EndOfBibitem
\bibitem[Wang \latin{et~al.}(2019)Wang, Zheng, Ye, Gan, Li, Song, Zhou, Ma, Yu,
  Gai, \latin{et~al.} others]{wang2019deep}
Wang,~M.; Zheng,~D.; Ye,~Z.; Gan,~Q.; Li,~M.; Song,~X.; Zhou,~J.; Ma,~C.;
  Yu,~L.; Gai,~Y., \latin{et~al.}  Deep graph library: A graph-centric,
  highly-performant package for graph neural networks. \emph{arXiv preprint
  arXiv:1909.01315} \textbf{2019}, \relax
\mciteBstWouldAddEndPunctfalse
\mciteSetBstMidEndSepPunct{\mcitedefaultmidpunct}
{}{\mcitedefaultseppunct}\relax
\EndOfBibitem
\bibitem[Pham and Wang(2015)Pham, and Wang]{pham2015oxygen}
Pham,~H.~H.; Wang,~L.-W. Oxygen vacancy and hole conduction in amorphous TiO 2.
  \emph{Physical Chemistry Chemical Physics} \textbf{2015}, \emph{17},
  541--550\relax
\mciteBstWouldAddEndPuncttrue
\mciteSetBstMidEndSepPunct{\mcitedefaultmidpunct}
{\mcitedefaultendpunct}{\mcitedefaultseppunct}\relax
\EndOfBibitem
\bibitem[Kumar \latin{et~al.}(2015)Kumar, Chernatynskiy, Liang, Choudhary,
  Noordhoek, Cheng, Phillpot, and Sinnott]{kumar2015charge}
Kumar,~A.; Chernatynskiy,~A.; Liang,~T.; Choudhary,~K.; Noordhoek,~M.~J.;
  Cheng,~Y.-T.; Phillpot,~S.~R.; Sinnott,~S.~B. Charge optimized many-body
  (COMB) potential for dynamical simulation of Ni--Al phases. \emph{Journal of
  Physics: Condensed Matter} \textbf{2015}, \emph{27}, 336302\relax
\mciteBstWouldAddEndPuncttrue
\mciteSetBstMidEndSepPunct{\mcitedefaultmidpunct}
{\mcitedefaultendpunct}{\mcitedefaultseppunct}\relax
\EndOfBibitem
\bibitem[Choudhary \latin{et~al.}(2015)Choudhary, Liang, Chernatynskiy,
  Phillpot, and Sinnott]{choudhary2015charge}
Choudhary,~K.; Liang,~T.; Chernatynskiy,~A.; Phillpot,~S.~R.; Sinnott,~S.~B.
  Charge optimized many-body (COMB) potential for Al2O3 materials, interfaces,
  and nanostructures. \emph{Journal of Physics: Condensed Matter}
  \textbf{2015}, \emph{27}, 305004\relax
\mciteBstWouldAddEndPuncttrue
\mciteSetBstMidEndSepPunct{\mcitedefaultmidpunct}
{\mcitedefaultendpunct}{\mcitedefaultseppunct}\relax
\EndOfBibitem
\bibitem[Guo \latin{et~al.}(2015)Guo, Liu, and Robertson]{guo2015chalcogen}
Guo,~Y.; Liu,~D.; Robertson,~J. Chalcogen vacancies in monolayer transition
  metal dichalcogenides and Fermi level pinning at contacts. \emph{Applied
  Physics Letters} \textbf{2015}, \emph{106}, 173106\relax
\mciteBstWouldAddEndPuncttrue
\mciteSetBstMidEndSepPunct{\mcitedefaultmidpunct}
{\mcitedefaultendpunct}{\mcitedefaultseppunct}\relax
\EndOfBibitem
\bibitem[Wu \latin{et~al.}(2020)Wu, Liu, Sun, Song, and Shi]{wu2020first}
Wu,~K.; Liu,~T.; Sun,~R.; Song,~J.; Shi,~C. First-principles calculations of
  oxygen vacancy in CaO crystal. \emph{The European Physical Journal D}
  \textbf{2020}, \emph{74}, 1--6\relax
\mciteBstWouldAddEndPuncttrue
\mciteSetBstMidEndSepPunct{\mcitedefaultmidpunct}
{\mcitedefaultendpunct}{\mcitedefaultseppunct}\relax
\EndOfBibitem
\end{mcitethebibliography}

\end{document}